\documentclass[mathleft
]{an}
\usepackage{graphicx}
\usepackage{times}
\usepackage{amsmath}
\DeclareMathOperator*{\med}{med}
\usepackage{natbib}

\renewcommand\refname{}

\begin{document}

\Pagespan{26}{29}
\Yearpublication{2012}%
\Yearsubmission{2011}%
\Month{09}%
\Volume{333}%
\Issue{1}%
\DOI{10.1002/asna.201111614}%

\title{A magnetic cycle of $\tau$~Bootis? The coronal and chromospheric view}

\author{K. Poppenhaeger\inst{1}\fnmsep\thanks{Corresponding author:
  {katja.poppenhaeger@hs.uni-hamburg.de}}
\and  H.M. G\"unther\inst{2}
\and J.H.M.M. Schmitt\inst{1}
}
\titlerunning{$\tau$~Boo's magnetic cycle}
\authorrunning{K. Poppenhaeger, H.M. G\"unther \& J.H.M.M. Schmitt}
\institute{
Hamburger Sternwarte, Gojenbergsweg 112, 21029 Hamburg, Germany
\and 
Harvard-Smithsonian Center for Astrophysics, 60 Garden Street, Cambridge, MA 02138, USA
}

\received{2011 Sep 19}
\accepted{2011 Oct 6}
\publonline{2012 Jan 23}


\abstract{%
$\tau$~Bootis is a late F-type main sequence star orbited by a Hot Jupiter. During the last years spectropolarimetric observations led to the hypothesis that this star may host a global magnetic field that switches its polarity once per year, indicating a very short activity cycle of only one year duration. In our ongoing observational campaign, we have collected several X-ray observations with XMM-Newton and optical spectra with TRES/FLWO in Arizona to characterize $\tau$~Boo's corona and chromosphere over the course of the supposed one-year cycle. Contrary to the spectropolarimetric reconstructions, our observations do not show indications for a short activity cycle.}

\maketitle

\section{Introduction}

Magnetic activity is an ubiquitous phenomenon in late-type stars and manifests itself in both short-term processes such as flares and coronal mass ejections as well as in long-term observables such as the solar eleven-year activity cycle. For a multitude of cool stars such cycles of several years duration have been found by monitoring their chromospheric activity indicators \citep{BaliunasDonahueSoon1995}; however, a thorough understanding of stellar activity cycles and their dependence on fundamental stellar parameters is still missing.
 
Especially for planet-hosting stars the stellar activity is highly interesting as high-energy radiation, stellar wind and coronal mass ejections are expected to have a strong influence on the outer planetary atmosphere. Planets have been detected around stars of very different activity levels. The first (radial-velocity) detected planet 51\,Peg\,b probably orbits a Maunder minimum star \citep{Poppenhaeger2009}, while especially space-based transit observations have also discovered planets around young and active stars like \hbox{Corot-2} \citep{Alonso2008}. In addition, several stars with indications for an activity cycle similar to the solar sunspot cycle have been found to host planets, such as $\iota$~Hor \citep{Metcalfe2010} and $\tau$~Boo, the latter of which is closely investigated for its coronal and chromospheric activity in this paper.

   \begin{table}
      \caption[]{{XMM-Newton} and optical observations of $\tau$~Boo with exposure time $t$
 given and expected activity state as extrapolated from magnetic field reconstructions.}
        \label{xmmobs}
\tabcolsep=9.5pt
    \begin{tabular}{l l r l}
    \hline\noalign{\smallskip}
    Obs. ID & Obs. Date & $t$ (ks) & State \\[1.5pt] \hline\noalign{\smallskip}
    0144570101 & 2003-06-24 & 70.5 & min. \\
    0651140201 & 2010-06-19 & 12.7 & min. \\ 
    (optical)  & 2010-06-19 & 2.2  & min. \\
    0651140301 & 2010-07-23 & 7.7  & min. \\ 
    (optical)  & 2010-07-24 & 1.7  & min. \\
    0651140401 & 2010-12-19 & 9.7  & max. \\
    0651140501 & 2011-01-22 & 13.3 & max. \\ 
    (optical)  & 2011-04-(07-18) & 6.2 & int.\\
    0671150501 & 2011-06-19 & 11.6 & min. \\ 
    (optical)  & 2011-07-(15-17) & 6.1 & min.\\[1.5pt] \hline
    \end{tabular}
   \end{table}

$\tau$~Boo is a planet-hosting main sequence star of spectral type F7 located at
$15.6$~pc distance from the Sun. Its age has been estimated to be roughly 3~Gyr
from isochrones, lithium abundances and chromospheric \ion{Ca}{ii} activity
\citep{Saffe2005}. For this age, the star rotates rather fast with a mean
rotation period of $P_\ast = 3.23$~d; it also displays quite strong differential
rotation with ${P_{\rm eq}=3}$~d and $P_{\rm pole}=3.9$~d at the equator and the poles, respectively \citep{Donati2008}. It has been speculated that this fast rotation stems from a tidal spin-up induced by the giant planet that orbits the star with a period of $3.3$~d \citep{Barnes2001}. 

Even if magnetic activity is not understood well enough to predict durations and
strengths of activity cycles from fundamental stellar parameters, a short
activity cycle might be expected for $\tau$~Boo as stellar rotation and magnetic
activity are related in late-type stars. In the Mount Wilson program
\citep{BaliunasDonahueSoon1995}, the star did not exhibit strong periodic
activity changes; some weak indications for a 12~yr periodicity were found, but were rated as ''poor'' in terms of false-alarm probability by the authors. During the last years, the
large-scale magnetic field of $\tau$~Boo was reconstructed from
spectropolarimetric measurements using Zeeman Doppler Imaging
\citep{Catala2007, Donati2008, Fares2009}. These reconstructions suggested that the polarity of the large-scale magnetic field switched twice during a period of two years, indicating an activity cycle of only one year duration. 

\phantom{ }

If these reconstructions really characterize the actual \linebreak magnetic field configuration of the star, it can be expected in analogy to the Sun that $\tau$\,Boo is in a state of minimum activity during the phases of a stable, poloidal field configuration. During the polarity switches, when toroidal field configurations are dominant, the activity level should be at a maximum. The available Zeeman Doppler Imaging (ZDI) data suggest that the polarity switches occur yearly in winter; our observations therefore cover several winter and \linebreak summer pointings to look for systematic changes.

\section{Observations and data analysis}\label{analysis}

\begin{figure}
\hskip-5mm
\includegraphics[width=0.51\textwidth]{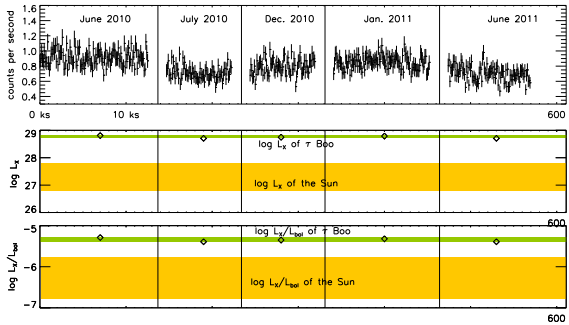}
\caption{(online colour at: www.an-journal.org) {\it Upper panel:}
background-subtracted X-ray lightcurves of $\tau$~Boo in 2003 and 2010/11 with
100~s time binning, observed with the {XMM-Newton} PN camera; {\it middle
panel:} range of $\log L_{\rm X}$ of $\tau$~Boo (green) in comparison to the range
covered by the Sun during an activity cycle (yellow); {\it lower panel:} same
for range of $\log (L_{\rm X}/L_{\rm bol})$.}
\label{xraylc}
\end{figure}

\subsection{X-ray data}

We monitored $\tau$\,Boo's X-ray emission with the  {XMM-Newton} telescope in five observations so far; see Table~\ref{xmmobs}. 
Additionally, there is an archival {XMM-Newton} observation of the star from June 2003. The data from this observation has been analyzed in detail by \cite{Maggio2011}; however, we have re-analyzed the dataset along the same lines as we have done for our new observations from  2010/11 for better comparability. All observations were performed with {XMM-Newton's} thick filter, as $\tau$~Boo is an optically bright target with $m_V=4.5$. This is also the reason why the optical monitor of {XMM-Newton} had to be blocked and could not be used for scientific analysis.

We reduced the data using standard procedures of the SAS10.0 software package. $\tau$~Boo has a mean X-ray countrate of $\approx$0.8~cts\,s$^{-1}$, practically all photons have energies below 5~keV, except for the observation in 2003 where also few X-ray source photons of higher energies were collected. We produced light curves with 100~s binning to obtain acceptable error bars as well as enough time resolution to identify possible flares. For the spectra, we used energy bins with at least 15 counts per bin for decent statistics. We extracted EPIC CCD spectra with moderate spectral resolution as well as high-resolution grating spectra from the two RGS instruments. Significant background signal was present for the 2003 observation, so in analyzing this exposure we used good time intervals with low background signal to extract the source spectra. The spectral fitting was performed with Xspec~v12.5.

$\tau$~Boo has a stellar companion at an angular distance of 2.8'' \citep{Patience2002} which is unresolved in the XMM-Newton observations. This companion, GJ~527\,B, is a low-mass main-sequence star of spectral type M2. 
The majority of early M dwarfs ($\approx$80\,\%) have luminosities below ${\log L_{\rm X}=27.5}$ \citep{Schmitt1995}. 
This amounts to a fraction of only $5\,\%$ of the detected X-ray flux of both $\tau$~Boo and GJ~527~B together, so that we can safely choose to neglect the contribution of the low-mass companion to the X-ray emission in our observations.

\subsection{Optical data from FLWO}

The Fred Lawrence Whipple Observatory in Arizona hosts the TRES spectrograph at its 1.5-m telescope. TRES is a cross-dispersed echelle spectrograph with a resolution of $\approx$20\,000--40\,000 (depending on the fiber used) in a bandpass covering 3900--9100\,\AA. For our observations the  medium fiber was used, yielding a spectral resolution of $\approx$30\,000. The raw spectra were flatfielded and the wavelength calibration was conducted through ThAr reference frames, using the TRES reduction pipeline.

Optical data is available for June and July 2010 as well as for April and July 2011; in the 2010 observations, a total observation duration of ca. 30 minutes was reached, split into several individual pointings. In 2011, the star was observed during three nights each, yielding a total exposure time of ca. 1.5~h.

\section{Results}

\subsection{X-ray lightcurves}

The X-ray lightcurves of $\tau$~Boo, collected in summer 2003, summer 2010, winter 2010/11, and summer 2011, are shown in Fig.~\ref{xraylc}. The lightcurves were extracted from the PN detector in the 0.2--5~keV energy band. The median countrate in the 2003 observation was higher than in any of the later observations with ca. 1.0~cts\,s$^{-1}$. The 2010/11 observations displayed median countrates of 0.90, 0.69, 0.79, 0.85, and 0.70~cts\,s$^{-1}$, respectively. All lightcurves display some short-term variability of 10--30\,\%. The 2003 observation exhibits several small flares, and also the 2010/11 observations show a few flare-like variations on a very low level. However, the flares are too small to allow a detailed loop analysis.

A convenient way to characterize the variability of the light curves with a single number is the Median Absolute Deviation (MAD). It is defined as the median of the absolute deviations from the median in each individual lightcurve, explicitly $MAD = \med(|X_i - \med(X_i)|)$, with $X_i$ being the measured count rates per time bin of the light curve. The MAD values for each observation are given in Table~\ref{lx}.

\begin{figure}
\hskip-1mm
\includegraphics[width=0.485\textwidth]{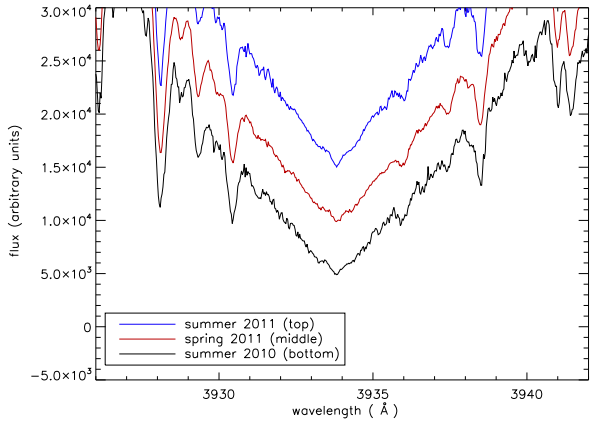}
\caption{(online colour at: www.an-journal.org) Ca\,{\sc ii} K line of $\tau$~Boo in summer 2010, spring 2011, and summer 2011; the spectra are vertically shifted for better visibility. The emission in the line core is practically unchanged.}
\label{opticalspectrum}
\end{figure}

\subsection{Activity levels}

A good indicator for coronal activity is the ratio of X-ray and bolometric
luminosity. Inactive stars typically display values of $\log (L_{\rm X}/L_{\rm bol}) < -6$; the Sun's activity index varies between $-6.8$ and $-5.8$ during an activity cycle \citep{Judge2003}.

   \begin{table}[t!]
      \caption[]{X-ray luminosity (0.2--10~keV), activity indicator $L_X/L_{\rm bol}$ and MAD (see text for explanation) during the five observations of $\tau$~Boo.}
        \label{lx}
\tabcolsep=8.5pt
    \begin{tabular}{l c c c c}
    \hline\noalign{\smallskip}
    Date 	& $L_{\rm X}$ 				& $\log
    \frac{L_{\rm X}}{L_{\rm bol}}$ 	& MAD 	& State	\\ 
		&(erg\,s$^{-1}$) 		&			&	&	\\[1.5pt] 
\hline\noalign{\smallskip}
     2003 Jun	& $7.6{\times} 10^{28}$			& --5.22   		& 0.090		& min.\\
     2010 Jun	& $6.5{\times} 10^{28}$			& --5.29  		& 0.072		& min.\\
     2010 Jul	& $5.1{\times} 10^{28}$			& --5.39   		& 0.050		& min.\\ 
     2010 Dec	& $5.6{\times} 10^{28}$			& --5.35   		& 0.072		& max.\\
     2011 Jan	& $6.1{\times }10^{28}$			& --5.32  		& 0.069		& max.\\ 
     2011 jun	& $5.1{\times} 10^{28}$			& --5.39			& 0.059		& min. \\[1.5pt]
    \hline
    \end{tabular}
   \end{table}

We compute the mean X-ray luminosity of $\tau$~Boo in each of the five observations by fitting MOS, PN and RGS spectra in Xspec 12.0 with a {VAPEC} model with four temperature components and variable abundances for the most prominent elements visible in the X-ray spectra, which are oxygen, neon, iron, magnesium, and silicon. We calculate the X-ray luminosity in the 0.2--10~keV energy band and the activity indicator from the spectral model, the results are given in Table~\ref{lx}. In cool stars, the activity indicator typically spans values of $-7$ to $-3$, placing $\tau$~Boo at a moderate level of activity which is higher than the solar activity level at the maximum of the solar cycle. The highest X-ray activity level was detected in summer 2003, where the X-ray luminosity was higher by 50\,\% compared to the lowest activity level detected in July 2010 and June 2011.

The chromospheric activity level can be determined \linebreak from the optical
spectra we recorded. The core of the \ion{Ca}{ii} K line, located at wavelengths
around $3933$\,\AA, is depicted in Fig.~\ref{opticalspectrum}. There is a small
amount of emission in the line core, typical for a low to moderate level of
activity. To quantify this emission, we calculate the equivalent width of the
\ion{Ca}{ii} K line core, contained in a $1$\,\AA~part of the spectrum centered
around the minimum of the core,  with respect to the pseudo-continuum present
between $3945$ and $3955$\,\AA. The values are very similar for the optical
observations with $EW_{\rm summer\,2010}=0.987$\,\AA, $EW_{\rm
spring\,2011}=0.989$\,\AA, and $EW_{\rm summer\,2011}=0.978$\,\AA. Further observations to be obtained in 2011/12 will give more insight into the variability of $\tau$~Boo's chromospheric activity.

\section{Discussion}

Our observations have shown that $\tau$~Boo is a moderately active star which displays some small-scale variability in X-rays. However, using the data available up to now, we do not find evidence for a short activity cycle of $\approx$1~yr duration. Especially an elevated activity state in winter 2010/11 as extrapolated from spectropolarimetric measurements is not present in the stellar coronal emission.

This is not a problem of identifying stellar activity cycles in X-ray emission. It has been shown for two stars other than the Sun, namely HD~81809 and 61~Cyg \citep{Favata2008, HempelmannRobrade2006}, that the quasi-quiescent coronal emission in general follows the chromospheric activity behavior. For these stars, the activity cycles with approximately 8 and 10~yr are much longer than the one that was proposed for $\tau$~Boo.

This leaves two main reasons why the coronal emission does not show the expected long-term variability. On the one hand, the sampling of our data available so far is quite sparse with only four pointings distributed over one year. It might be that we incidentally caught $\tau$~Boo in short phases of low activity during winter 2010/11, while the general activity level during that period was significantly higher. For the observations from 2003, a low activity state was extrapolated from the spectropolarimetric data. If truly a 1-year cycle is present, then there are seven cycles between that dataset and the 2010/11 observations. We know from the Sun that different activity cycles can be more or less pronounced, so the higher activity level in 2003 does not necessarily contradict this interpretation.

On the other hand, the magnetic polarity switches reconstructed from
spectropolarimetric measurements might not be caused by a short magnetic cycle
in the first place. In those observations, the Stokes $I$ and $V$ components were
measured, and the magnetic field reconstructions then yield mainly information
on the {\em net} magnetic field of the stellar hemisphere that is visible during
the individual observations. Areas on the stellar surface which have opposite
polarity ``cancel out" in the Stokes $V$ signal and can therefore usually not be reconstructed by measuring only these two components. If these areas and their magnetic fields differ slightly, the Stokes $V$ signature appears as that of the net field strength of both areas, and thus does not allow a distinction between global net fields and a locally differing field strength of opposite polarity. 

In the case of $\tau$~Boo, a net radial magnetic field with a strength of up to $10$~G has been reconstructed \citep{Fares2009}. In the Sun, the magnetic field strength in sunspots is of the order of several kilogauss, while the global polar field of the Sun is much weaker with only a few Gauss. Even if sunspots usually are present in pairs, it is well possible that a snapshot of one stellar hemisphere of $\tau$~Boo contains local magnetic fields in such a way that their integral over the stellar disk yields a net field strength equals $10$~G. 

\section{Conclusion}
In our ongoing observational campaign we monitor the \linebreak coronal and chromospheric activity of $\tau$~Boo to test for a short activity cycle as predicted from spectropolarimetric observations. We have collected X-ray and optical data over a period of one and a half years which show that $\tau$~Boo is a star of moderate and slightly variable activity. However, our data do not show indications for an activity cycle of one year duration.

\bibliographystyle{aa}
\bibliography{../../Docs/katjasbib.bib}

\end{document}